\documentclass[11pt]{revtex4}

\usepackage[cp1251]{inputenc}

\usepackage{amscd}

\usepackage{amsmath}

\begin{document}




\title{Quantum Kepler problem for spin 1/2 particle\\
in spaces on constant curvature. I. Pauli theory.  }

\medskip
\author{E.M. Ovsiyuk
          \\
              e.ovsiyuk@mail.ru
}

\begin{abstract}

Transition to a nonrelativistic Pauli equation in  Riemann
space of  constant positive curvature  for a Dirac  particle   in
presence of the Coulomb field is performed in the system of radial
equations, exact solutions  are constructed in  terms of Hein  functions, the
energy spectrum is derived.
The same is done for the Kepler quantum  problem in hyperbolic Lobachevsky space,
 solutions  are constructed in  terms of Hein
functions, the energy spectrum is obtained.


\end{abstract}

\maketitle

\subsection*{ Introduction}

Quantum mechanics had been started with the theory of the  hydrogen atom,
so  considering  the Quantum mechanics in Riemannian spaces it is first natural step to turn to just this system.
A common  quantum-mechanical hydrogen atom model is based materially
on  the assumption of the Euclidean character of the physical 3-space geometry.
In this context,  natural questions arise:  what in  such a model description is determined
by this assumption, and which changes  will be entailed by allowing for  other spatial geometries: for instance.
Lobachevsky's  $H_{3}$, Riemann's $S_{3}$, or de Sitter geometry. The question
is of fundamental significance, even beyond  its possible  experimental testing.

Firstly, the  hydrogen atom in 3-dimensional  space of constant positive curvature $S_{3}$  was considered by
Schr\"{o}dinger    \cite{1940-Schrodinger}. He had been studied the so-called factorization method in quantum mechanics; in particular,
application of this techniques to discrete  part of the energy spectrum for hydrogen atom had been elaborated.
An idea was to modify the basic  atom system so as to cover all the energy spectrum including the region $E>0$ as well.
However, mere placing of the atom system inside a finite box  in order  to make  the  whole energy spectrum discrete
 did not seem attractive, so  Schr\"{o}dinger had placed the atom into the curved background of the Riemann
 space model  $S_{3}$. Due its compactness, the spherical Riemann model may simulate the effect of the finite box
 -- see Schrodinger \cite{1940-Schrodinger} and  Stevenson \cite{1941-Stevenson}.

In spherical coordinates of $S_{3}$
\begin{eqnarray}
dl^{2} = d \chi^{2} + \sin^{2}\chi (d\theta^{2} + \sin^{2}d\phi^{2}) \;
\nonumber
\end{eqnarray}

\noindent the Shchr\'{o}dinger Hamiltonian in dimensionless units has the form
\begin{eqnarray}
H = -{1 \over 2} \; {1 \over \sqrt{g}} \; {\partial \over \partial
x^{\alpha}} \; \sqrt{g} \; g^{\alpha \beta}\; {\partial \over
\partial x^{\alpha}} \; -\; {e \over \mbox{tan}\;  \chi }  \;  ;
\nonumber
\end{eqnarray}

\noindent $\rho$ is a curvature radius, a unite for length;
$M$ is a mass of the electron; $\hbar^{2}
/M\rho^{2}$ is taken as a unit for energy;  $e = {\alpha \over \rho
} /  { \hbar^{2} \over M \rho^{2} } $ stands for a Coulomb interaction constant;
the sign at  $e / \mbox{tan}\; \chi $  corresponds to the attracting Coulomb force.
The energy spectrum is entire discrete and given by
\begin{eqnarray}
\epsilon_{n} = - {e^{2} \over 2 n^{2} } + {1 \over 2} \; (n^{2} -1) \; ,
 \;\;  n = 1,2, 3, ...
\nonumber
\end{eqnarray}

Hydrogen atom in the Lobachevsky space  $H_{3}$  was considered firstly by
 Infeld and Shild  \cite{1945-Infeld-Schild}
\begin{eqnarray}
dl^{2} = d \chi^{2} + \mbox{cosh}^{2}\chi (d\theta^{2} +
\sin^{2}d\phi^{2}) \;  ,
\nonumber
\\
H = -{1 \over 2}  {1 \over \sqrt{g}}  {\partial \over \partial
x^{\alpha}} \; \sqrt{g} \; g^{\alpha \beta}  {\partial \over
\partial x^{\alpha}}  -  {e \over \mbox{tanh}\;  \chi }  \; .
\nonumber
\end{eqnarray}

\noindent Energy spectrum contains a discrete and continuous parts. The number of discrete levels
is finite,  they are specified  by
\begin{eqnarray}
- \; {e^{2} \over 2} \leq \epsilon \leq ({1\over 2}  - e ) \; ,
\nonumber
\\
\epsilon_{n} = - {e^{2} \over 2 n^{2} } - {1 \over 2} \; (n^{2} -1) \; ,\;\;
    n = 1,2, 3, ...,N \; .
\nonumber
\end{eqnarray}

\noindent   In the region  $\epsilon \geq ({1\over2} - e)$ the energy spectrum is continuous.

Thus, the models of the hydrogen atom in Euclid, Riemann, and Lobachevsky spaces
significantly differ from each other,
 which is the result of differences in three spatial  geometries:
  $E_{3},\; H_{3},\; S_{3}$. To present time, we see a plenty of investigations
  on this matter:

Higgs \cite{1979-Higgs}, Leemon \cite{1979-Leemon}, Kurochkin --
Otchik \cite{1979-Kurochkin-Otchik}, Bogush -- Kurochkin -- Otchik
\cite{1980-Bogush-Kurochkin-Otchik}, Parker \cite{1980-Parker},
\cite{1981-Parker}, Ringwood -- Devreese
\cite{1980-Ringwood-Devreese}, Kobayshi \cite{1980-Kozo}, Bessis
-- Bessis -- Shamseddine \cite{1982-Bessis-Bessis-Shamseddine},
Grinberg -- Maranon -- Vucetich
\cite{1983-Grinberg-Maranon-Vucetich}, Bogush -- Otchik -- Red'kov
\cite{1983-Bogush-Otchik-Red'kov}, Bessis -- Bessis -- Shamseddine
\cite{1984-Bessis-Bessis-Shamseddine(1)},\cite{1984-Bessis-Bessis-Roux(2)},
\cite{1984-Bessis-Bessis(3)}, Chondming -- Dianyan
\cite{1984-Chondming-Dianyan}, Xu -- Xu \cite{1984-Xu-Xu},
Melnikov -- Shikin \cite{1985-Melnikov-Shikin}, Shamseddine
\cite{1986-Shamseddine}, Otchik -- Red'kov
\cite{1986-Otchik-Red'kov}, Barut -- Inomata -- Junker
\cite{1987-Barut-Inomata-Junker}, Bessis -- Bessis -- Roux
\cite{1988-Bessis-Bessis-Roux}, Bogush -- Otchik -- Red'kov
\cite{1988-Bogush-Otchik-Red'kov.}, Gorbatsievich -- Priebe
\cite{1989-Gorbatsievich-Priebe}, Groshe \cite{1990-Groshe}, Barut
-- Inomata -- Junker \cite{1990-Barut-Inomata-Junker}, Katayama
\cite{1990-Katayama}, Chernikov \cite{1992-Chernikov}, Mardoyan --
Sisakyan \cite{1992-Mardoyan-Sisakyan}, Granovskii -- Zhedanov --
Lutsenko \cite{1992-Granovskii-Zhedanov-Lutsenko}, Kozlov -- Harin
\cite{1992-Kozlov-Harin}, Vinitskii -- Marfoyan -- Pogosyan --
Sisakyan -- Strizh
\cite{1993-Vinitskii-Marfoyan-Pogosyan-Sisakyan-Strizh},
Shamseddine \cite{1997-Shamseddine}, Bogush -- Kurochkin -- Otchik
\cite{1998-Bogush-Kurochkin-Otchik}, Otchik \cite{1999-Otchik},
Nersessian -- Pogosyan \cite{2001-Nersessian-Pogosyan}, Red'kov
\cite{2001-Red'kov}, Bogush -- Kurochkin -- Otchik
\cite{2003-Bogush-Kurochkin-Otchik}, Kurochkin -- Otchik --
Shoukavy \cite{2006-Kurochkin-Otchik-Shoukavy}, Kurochkin --
Shoukavy \cite{2006-Yu. Kurochkin-Shoukavy}, Bogush -- Otchik --
Red'kov \cite{2006-Bogush-Otchik-Red'kov}, Bessis -- Bessis
\cite{1979-Bessis-Bessis},  Iwai \cite{1982-Iwai}, Cohen -- Powers
\cite{1982-Cohen-Powers}.

\subsection*{1. Separation of the variables  for Dirac equation in  curved models}

Let us consider procedure of separation of the variables in  the Dirac equation on the background of
curved model, fir definiteness  using a spherical Riemann model
  $S_{3}$, a diagonal tetrad is taken in the form
\begin{eqnarray}
dS^{2} = dt^{2} - d \chi^{2} - \sin^{2} \chi ( d \theta^{2} + \sin^{2} d \phi^{2}) \; ,
\nonumber
\\
e_{(0)}^{\alpha} = ( 1,0,0,0) \; , \qquad
e_{(3)}^{\alpha} = ( 0, 1,0,0) \; , \;
\nonumber
\\
e_{(1)}^{\alpha} = ( 0,0, \sin^{-1} \chi, 0) \; , \;
e_{(2)}^{\alpha} = ( 0,0, 0, \sin^{-1} \chi \sin^{-1} \theta ) \;
.
\label{1.1}
\end{eqnarray}

\noindent
Generally covariant Dirac equation
\begin{eqnarray}
\left [  i \gamma^{c} ( e_{(c)}^{\alpha} \partial _{\alpha} +
{1 \over 2} j^{ab} \gamma_{abc}) -m  \right  ] \; \Psi = 0
\nonumber
\end{eqnarray}
takes the form
\begin{eqnarray}
\left [ \; i \gamma^{0} {\partial \over \partial t } \;  + \; i
(\gamma^{3} {\partial \over \partial \chi} + {\gamma^{1} j^{31} +
\gamma^{2} j^{32} \over \mbox{tan}\;  \chi )}) \; + \; {1 \over
\sin \chi } \Sigma_{\theta \phi}\; -\;  m \right  ] \; \Psi = 0
\;,
  \label{1.2a}
\end{eqnarray}

\noindent where an angular operator $\Sigma _{\theta ,\phi }$ is
defined by
\begin{eqnarray}
\Sigma _{\theta ,\phi } =  \; i\;
\gamma ^{1}
\partial _{\theta } \;+\; \gamma ^{2}\; {i\partial \;+\;i\;j^{12}\cos
\theta  \over \sin \theta  } \; .
  \label{1.2a}
\end{eqnarray}

\noindent
Allowing for
\begin{eqnarray}
{\gamma^{1} j^{31} + \gamma^{2} j^{32} \over \mbox{tan}\; \chi} = { \gamma^{3} \over
\mbox{tan}\;  \chi  } \; , \qquad  \Psi = {1 \over \sin \chi } \tilde{\Psi}
\nonumber
\end{eqnarray}

\noindent  eq. (\ref{1.2a}) is simplified
\begin{eqnarray}
\left [  i \gamma^{0} {\partial \over \partial t }   +  i
\gamma^{3} {\partial \over \partial \chi}  +  {1 \over \sin
\chi } \Sigma_{\theta \phi} -  m   \right ]  \tilde{\Psi} = 0 \; .
\label{1.2b}
\end{eqnarray}

\noindent To diagonalize operators  $i\partial_{t}, \vec{J}^{2}, J_{3}$,
one takes the wave function in the form  \cite{1988-Red'kov},
\cite{1998-Red'kov}
\begin{eqnarray}
\tilde{\Psi} = e^{-i \epsilon t }
\left | \begin{array}{r}
f_{1}(\chi) \; D_{-1/2} \\
f_{2}(\chi) \; D_{+1/2} \\
f_{3}(\chi) \; D_{-1/2} \\
f_{4}(\chi) \; D_{+1/2}
\end{array} \right | \; ;
\label{1.3}
\end{eqnarray}

\noindent where Wigner functions   are noted as $D_{\sigma}= D^{j}_{-m.\sigma}(\phi, \theta,0)$.
After separation of the variables we get four radial equations
(let   $ \nu =j+1/2$) :
\begin{eqnarray}
\epsilon   f_{3}   -  i  {d \over d \chi}  f_{3}   -
i {\nu \over \sin \chi }  f_{4}  -  m  f_{1} =   0  \; ,
\qquad \epsilon   f_{4}   +  i  {d \over d \chi } f_{4}   +
i {\nu \over \sin \chi}  f_{3}  -  m  f_{2} =   0    \;  ,
\nonumber
\\
\epsilon   f_{1}   +  i  {d \over d \chi }  f_{1}  + i {\nu \over
\sin \chi }  f_{2}  -  m  f_{3} =   0  \; ,
\qquad
 \epsilon
f_{2}   -  i  {d \over d \chi } f_{2}   - i {\nu \over \sin \chi }
f_{1}  -  m  f_{4} =   0  \; .
\label{1.4}
\end{eqnarray}

\noindent In spherical tetrad, the  space reflection  operator is given by
\begin{eqnarray}
\hat{\Pi}_{sph.} \; \; =
\left | \begin{array}{cccc}
0 &  0 &  0 & -1   \\
0 &  0 & -1 &  0   \\
0 &  -1&  0 &  0   \\
-1&  0 &  0 &  0
\end{array} \right |
\; \otimes  \; \hat{P} \; .
\nonumber
\end{eqnarray}

\noindent From eigenvalues equations   $\hat{\Pi}_{sph.}\; \Psi _{jm}
= \; \Pi \; \Psi _{jm}$
we obtain
\begin{eqnarray}
\Pi = \; \delta \;  (-1)^{j+1} , \;\; \delta  = \pm 1 \; , \qquad
f_{4} = \; \delta \;  f_{1} , \qquad  f_{3} = \;\delta \; f_{2}
\; , \label{1.5}
\end{eqnarray}

\noindent which simplifies  (\ref{1.4}):
\begin{eqnarray}
\epsilon   f_{1}   +  i  {d \over d \chi }  f_{1}  + i {\nu \over
\sin \chi }  f_{2}  -  \delta m  f_{2} =   0  \; ,
\nonumber
\\
 \epsilon
f_{2}   -  i  {d \over d \chi } f_{2}   - i {\nu \over \sin \chi }
f_{1}  -  \delta m  f_{1} =   0  \; .
\label{1.6}
\end{eqnarray}

\noindent In terms of new functions
\begin{eqnarray}
f  =  {f_{1} + f_{2} \over \sqrt{2}} \; , \qquad
g  =  {f_{1} - f_{2} \over i \sqrt{2}} \; ,
\nonumber
\end{eqnarray}

\noindent the above system reads
\begin{eqnarray}
({d \over d \chi }  + {\nu \over \sin \chi } ) \; f   +   ( \epsilon  +
 \delta   m )\; g  = 0 \; ,\qquad
({d \over d \chi }   -  {\nu \over \sin \chi } )\; g   -   ( \epsilon  -
 \delta   m ) \; f =  0     \; .
\label{1.7}
\end{eqnarray}

\subsection*{2.  Pauli equation for Kepler problem, flat Minkowski space}

Let us  consider the problem of spinor spherical waves in Pauli
approximation. It is convenient to start with the radial  system
for a free particle case. As a first step, one should separate
the rest energy --  for this  it is enough to make  a  formal replacement
 $\epsilon \Longrightarrow \epsilon + m$:
\begin{eqnarray}
 ({d \over dr} \;+\; {\nu \over
r}\;) \; f \; + \; ( \epsilon  + m \;+ \;
 \delta \; m )\; g \; = \;0 \; ,
 \nonumber
\\
 ({d \over dr} \; - \;{\nu \over r}\;)\; g  \;- \; ( \epsilon  + m
\; - \;
 \delta\;  m )\; f\; =\; 0     \; .
\label{2.1}
\end{eqnarray}

\noindent States with opposite parity are specified by
\begin{eqnarray}
\delta =+1\;, \qquad ({d \over dr} \;+\; {\nu \over r}\;) \; f \;
+ \; ( \epsilon  + 2 m )\; g \; = \;0 \; , \;\;
 ({d \over dr} \;
- \;{\nu \over r}\;)\; g  -   \epsilon  \; f =\; 0     \; ;
\label{2.2a}
\end{eqnarray}
\begin{eqnarray}
\delta =-1\;, \qquad ({d \over dr} \;+\; {\nu \over r}\;) \; f \;
+ \;
 \epsilon  \; g \; = \;0 \; ,\;\;
({d \over dr} \; - \;{\nu \over r}\;)\; g  \;- \; ( \epsilon  + 2m
)\; f =\; 0     \; .
\label{2.2b}
\end{eqnarray}

\noindent With additional assumption  $ \epsilon  + 2 m \approx 2m
$, in each case one gets
 radial Pauli equation for  a big  component:
\begin{eqnarray}
\delta =+1\;, \qquad f >> g \; ,\qquad ( {d^{2} \over dr^{2}} +
{\nu^{2} + \nu \over r^{2}} )\;f + 2m \epsilon \; f=0\; ;
\nonumber
\\
\delta =-1\;, \qquad g >> f \; , \qquad ( {d^{2} \over dr^{2}} +
{\nu^{2} -\nu \over r^{2}})\;g+ 2 m \epsilon\; g =0\;. \label{2.3}
\end{eqnarray}

\noindent Corresponding 2-component nonrelativistic spherical
functions with different parity are
\begin{eqnarray}
\psi _{jm,\delta=+1}={ e^{i\epsilon t} \over r}\; \left |
\begin{array}{c}
f(r) \; D_{-1/2}\\
f(r) \; D_{+1/2}
\end{array} \right |\;, \qquad
\psi _{jm,\delta=-1}={ e^{i\epsilon t} \over r}\; \left |
\begin{array}{c}
ig\;(r) \; D_{-1/2}\\
-ig\;(r) \; D_{+1/2}
\end{array} \right |\;.
\label{2.4}
\end{eqnarray}

To obtain equation in presence of  the Coulomb field, it is enough
to make a formal replacement $\epsilon \Longrightarrow \epsilon
+\alpha / r$ in (\ref{2.1}):
\begin{eqnarray}
 ({d \over dr} \;+\; {\nu \over
r}\;) \; f \; + \; ( \epsilon  + {\alpha \over  r} \;+ \;
 \delta \; m )\; g \; = \;0 \; ,
 \nonumber
\\
 ({d \over dr} \; - \;{\nu \over r}\;)\; g  \;- \; ( \epsilon + {\alpha \over r}
\; - \;
 \delta\;  m )\; f\; =\; 0     \; .
\label{2.5}
\end{eqnarray}

After separating the rest energy, for states with different
parities we get
\begin{eqnarray}
\delta =+1\;, \qquad ({d \over dr} \;+\; {\nu \over r}\;) \; f \;
+ \; (  {\alpha \over r }  + 2 m )\; g \; = \;0 \; ,
\nonumber
\\
\qquad \qquad ({d \over dr} \; - \;{\nu \over r}\;)\; g  -  (
\epsilon   + {\alpha \over r } )\; f =\; 0     \; ; \label{2.6a}
\end{eqnarray}
\begin{eqnarray}
\delta =-1\;, \qquad ({d \over dr} \;+\; {\nu \over r}\;) \; f \;
+ \;
 (\epsilon  + {\alpha \over r}) \; g \; = \;0 \; ,
\nonumber
\\
\;\; \qquad \qquad  \qquad ({d \over dr} \; - \;{\nu \over r}\;)\;
g  \;- \; ( {\alpha \over r}  + 2m )\; f =\; 0     \; .
\label{2.6b}
\end{eqnarray}

 To derive the  Pauli equation with known structure,
 one must impose additional restriction
\begin{eqnarray}
 \epsilon  + {\alpha \over r} + 2 m \approx 2m
\nonumber
\end{eqnarray}

 \noindent  which means in fact that Pauli description cannot be goof enough
 in  the region close to the origin $r=0$ where a source of the Coulomb  field is located.
Thus, we obtain
\begin{eqnarray}
\delta =+1\;, \qquad ({d \over dr} \;+\; {\nu \over r}\;) \; f \;
+ \;  2 m \; g \; = \;0 \; , \qquad ({d \over dr} \; - \;{\nu
\over r}\;)\; g  -  ( \epsilon   + {\alpha \over r } )\; f =\; 0
\; ;
\label{2.7a}
\end{eqnarray}
\begin{eqnarray}
\delta =-1\;, \qquad ({d \over dr} \;+\; {\nu \over r}\;) \; f \;
+ \;
 (\epsilon  + {\alpha \over r}) \; g \; = \;0 \; ,
  \qquad ({d \over dr} \; - \;{\nu \over r}\;)\; g  \;- \;  2m  \; f =\; 0     \; .
\label{2.7b}
\end{eqnarray}

\noindent Respective radial equations  for big components are
\begin{eqnarray}
\delta =+1\;, \qquad  {d^{2}f\over dr^{2}}-\left({\nu(\nu+1)\over
r^{2}}-{2m\alpha\over r}-2m\epsilon\right)f=0 \; ; \label{2.8a}
\\
\delta =-1\;, \qquad  {d^{2}g\over dr^{2}}-\left({\nu(\nu-1)\over
r^{2}}-{2m\alpha\over r}-2m\epsilon\right)g=0 \; ; \label{2.8b}
\end{eqnarray}

\noindent two last equations can be related by the formal change $\nu
\Longrightarrow \nu -1$.

From (\ref{2.8a}), introducing a new variable
$x=2\sqrt{-2m\epsilon}\,r$, one gets
\begin{eqnarray}
x{d^{2}f\over dx^{2}}- \left( \; {\nu(\nu+1)\over x}+{x\over 4} -
\alpha \sqrt{-{m\over 2\epsilon}} \; \right)f=0 \; . \label{2.9a}
\end{eqnarray}

\noindent Searching solutions in the form  $f=x^{A}e^{-Cx}\; F$,
one derives
\begin{eqnarray}
x{d^{2}F\over dx^{2}}+(2A-2Cx){dF\over dx}+\left({A(A-1)\over x}
-2AC+C^{2}x-{\nu(\nu+1)\over x}-{x\over 4} +  \alpha \sqrt{-
{m\over 2\epsilon}}\right)F=0\,.
\nonumber
\end{eqnarray}

\noindent At special choice of  $A$ and $C$  (underlined values
correspond to  bound states)
\begin{eqnarray}
A=-\nu\,,\;\underline{1+\nu}\,,\qquad C= -1/2, \underline{+ 1/2 }
\nonumber
\end{eqnarray}

\noindent we simplify the problem
\begin{eqnarray}
x{d^{2}F\over dx^{2}}+(2A-x){dF\over dx}-\left(A - \alpha
\sqrt{-{m\over 2\epsilon}}\right)F=0\,. \label{2.10}
\end{eqnarray}

\noindent It is a confluent hypergeometric equation  with
parameters
\begin{eqnarray}
a=A - \alpha \sqrt{-{m\over 2\epsilon}}\,,\qquad c=2A\,.
\nonumber
\end{eqnarray}

\noindent Making series a  polynomial in usual way: $ a=-n\,,\;\;
n=0,1,2... $,  one  obtains  the energy quantization rule
(remembering  $\nu = j +1/2$)
\begin{eqnarray}
1 + \nu - \alpha \sqrt{-{m\over 2\epsilon}}=-n \qquad
\Longrightarrow \qquad \epsilon = -{m\alpha^{2} \over  2 (n + \nu
+1)^{2} } \; . \label{2.11}
\end{eqnarray}

Turning to eq. (\ref{2.8b}), by means of  the formal replacement $\nu
\Longrightarrow \nu -1$, we get
\begin{eqnarray}
\nu - \alpha \sqrt{-{m\over 2\epsilon}}=-n \qquad \Longrightarrow
\qquad \epsilon = -{m\alpha^{2} \over  2 (n + \nu )^{2} } \; .
\label{2.12}
\end{eqnarray}

\subsection*{3.  Pauli equation for  Kepler problem in spherical  space}

In the spherical model, again let  us start with  free
radial equations (compare with (\ref{2.2a}), (\ref{2.2b}))
\begin{eqnarray}
\delta =+1\;, \qquad ({d \over d \chi } \;+\; {\nu \over \sin \chi
}\;) \; f \; + \; 2 m \; g \; = \;0 \; ,
 \qquad ({d \over d \chi } \;
- \;{\nu \over \sin \chi }\;)\; g  -   \epsilon  \; f =\; 0     \;
;
\label{3.1a}
\\
\delta =-1\;, \qquad ({d \over d \chi } \;+\; {\nu \over \sin \chi
}\;) \; f \; + \;
 \epsilon  \; g \; = \;0 \; ,
\qquad ({d \over d \chi } \; - \;{\nu \over \sin \chi }\;)\; g
\;- \;  2m \; f =\; 0     \; .
\label{3.1b}
\end{eqnarray}

\noindent Correspondingly, in each case the Pauli radial equation
for a big component is
\begin{eqnarray}
\delta =+1\;, \qquad f >> g \; ,\qquad  {d^{2}f\over
d\chi^{2}}-\left({\nu(\nu+\cos \chi)\over \sin^{2} \chi}-2\epsilon
m\right) \; f=0\; ; \label{3.2a}
\\
\delta =-1\;, \qquad g >> f \; , \qquad  {d^{2}g\over
d\chi^{2}}-\left({\nu(\nu-\cos \chi)\over \sin^{2} \chi}-2\epsilon
m\right) \; g =0\;. \label{3.2b}
\end{eqnarray}

After  changing  the variable in  (\ref{3.2a}), $ y=(1+\cos x)/2$, eq.  (\ref{3.2a}) reads
\begin{eqnarray}
y\,(1-y)\,{d^{2}f\over dy^{2}}+({1\over2}-y)\,{df\over
dy}+\left[-{1\over 4}\,{\nu\,(\nu+1)\over 1-y}-{1\over
4}\,{\nu\,(\nu-1)\over y}+2\epsilon m\right]f=0\,.
\nonumber
\end{eqnarray}

\noindent Searching solution in the form $f=y^{A}\,(1-y)^{B}F$,
 for  $F$ we obtain
\begin{eqnarray}
y\,(1-y)\,{d^{2}F\over dy^{2}}+\left[2A+{1\over
2}-(2A+2B+1)y\right]\,{dF\over dy}+
\nonumber
\\
+\left[{A(A-1/2)-1/4\nu(\nu-1)\over
y}+{B(B-1/2)-1/4\nu(\nu+1)\over 1-y}-(A+B)^{2}+2\epsilon
m\right]F=0\,.
\nonumber
\end{eqnarray}

\noindent At  $A,\; B$ taken according to
\begin{eqnarray}
A=  \underline{\nu/ 2 } \,,\;(-\nu+1)/ 2\,,\qquad
 B=- \nu / 2 \,,\; \underline{ (\nu+ 1 )/ 2 }\,,
\nonumber
\end{eqnarray}

\noindent we  get to  a hypergeometric equation
\begin{eqnarray}
y\,(1-y)\,{d^{2}F\over dy^{2}}+\left[2A+{1\over
2}-(2A+2B+1)y\right]\,{dF\over dy} -\left[(A+B)^{2}-2\epsilon
m\right]F=0
\nonumber
\end{eqnarray}

\noindent  with parameters given by
\begin{eqnarray}
\alpha=A+B+\sqrt{2 \epsilon m}\,,\qquad \beta= A+B-\sqrt{2
\epsilon m}\,, \qquad \gamma=2A+{1\over 2}\,.\label{3.3}
\end{eqnarray}

\noindent Bound states are separated by
\begin{eqnarray}
A= \nu / 2 \,,\qquad B= (\nu+1)/ 2 \,,
\nonumber
\\
\beta = \nu+{1\over 2} - \sqrt{2 \epsilon m}=-n \; , \qquad
\epsilon= + \; {(n+\nu+1/2)^{2}\over 2m}\,. \label{3.4}
\end{eqnarray}

To treat eq. (\ref{3.2b}), it suffices to make a formal change $\nu
\Longrightarrow - \nu$,  so we arrive ar
\begin{eqnarray}
A=  -\nu/ 2  \,,\;(\nu+1)/ 2\,,\qquad
 B= \nu / 2 \,,\;  (-\nu+ 1 )/ 2 ,
\nonumber
\end{eqnarray}

\noindent and further
\begin{eqnarray}
y\,(1-y)\,{d^{2}F\over dy^{2}}+\left[2A+{1\over
2}-(2A+2B+1)y\right]\,{dF\over dy} -\left[(A+B)^{2}-2\epsilon
m\right]F=0\,
\nonumber
\end{eqnarray}

\noindent  with parameters
\begin{eqnarray}
\alpha=A+B+\sqrt{2 \epsilon m}\,,\qquad \beta= A+B-\sqrt{2
\epsilon m}\,, \qquad \gamma=2A+{1\over 2}\,. \label{3.5}
\end{eqnarray}

Bound states are  specified by
\begin{eqnarray}
A= (\nu+1)/2  \,,\qquad B= \nu / 2 \,,
\nonumber
\\
\beta = \nu+{1\over 2} - \sqrt{2 \epsilon m}=-n \; , \qquad
 \epsilon= + \; {(n+\nu+1/2)^{2}\over 2m}\,.
\label{3.6}
\end{eqnarray}

\noindent
Corresponding 2-component wave functions with opposite parities
are  given by
\begin{eqnarray}
\psi _{jm,\delta=+1}={ e^{i\epsilon t} \over \sin \chi}\; \left |
\begin{array}{r}
f(\chi) \; D_{-1/2}\\
f(\chi) \; D_{+1/2}
\end{array} \right |\;,\qquad
\psi _{jm,\delta=-1}={ e^{i\epsilon t} \over \sin \chi}\; \left |
\begin{array}{r}
ig\;(\chi) \; D_{-1/2}\\
-ig\;(\chi) \; D_{+1/2}
\end{array} \right |\;.
\label{3.7}
\end{eqnarray}

Now let us add the Coulomb potential
\begin{eqnarray}
\delta =+1\;, \qquad ({d \over d \chi } + {\nu \over \sin \chi
})  f  +  2 m  g  = 0 \; ,\;\;
 ({d \over d \chi }  - {\nu \over
\sin \chi } ) g  -   (\epsilon  + {\alpha \over  \mbox{tan}\;
\chi} )  f = 0     \; ;
\label{3.8a}
\\
\delta =-1\;, \qquad ({d \over d \chi } + {\nu \over \sin \chi
})  f  +
 (\epsilon   + {\alpha \over  \mbox{tan}\; \chi})  g  = 0 \; ,\;\;
 ({d \over d \chi }  -  {\nu \over \sin
\chi } ) g  -   2m  f = 0     \; .
\label{3.8b}
\end{eqnarray}

\noindent Respective   Pauli radial equations are
\begin{eqnarray}
\delta =+1\;, \qquad
 {d^{2}f \over d \chi^{2}}-\left(
 {\nu(\nu+\cos \chi)\over \sin^{2} \chi}- 2\epsilon m- {2 m \alpha \over \tan \chi}\right) \; f=0\,,
\label{3.9a}
\\
\delta =-1\;, \qquad
 {d^{2}g\over d\chi^{2}}-
 \left({\nu(\nu-\cos \chi)\over \sin^{2} \chi}-2\epsilon m-{2 m \alpha \over \tan \chi}\right) \; g =0 \; .
\label{3.9b}
\end{eqnarray}

Behavior of the function from  (\ref{3.9a}) near the points $\chi=0,\;
\pi$ is characterized by
\begin{eqnarray}
\chi \sim 0, \qquad {d^{2}f\over d\chi^{2}}-  {\nu(\nu+ 1)\over
\sin^{2} \chi}  \; f=0\,,
\nonumber
\\
 f = \sin^{A} \chi \; , \qquad
A= \underline{1+\nu},\;\; - \nu \; ;
\nonumber
\\
\chi \sim  \pi =  (\pi - \beta), \qquad {d^{2}f\over d\chi^{2}}-
{\nu(\nu - 1)\over \sin^{2} \chi}  \; f=0\,,
\nonumber
\\
 f = \sin^{B} (\pi - \beta)  \; , \qquad  B = \underline{+ \nu},\;\; 1 - \nu \; ,
\label{3.10a}
\end{eqnarray}

\noindent and in the case  (\ref{3.9b})
\begin{eqnarray}
\chi \sim 0, \qquad {d^{2}g\over d\chi^{2}}-  {\nu(\nu- 1)\over
\sin^{2} \chi}  \; g=0\,,
\nonumber
\\
 g = \sin^{A} \chi \, , \qquad
A= \underline{+ \nu},\;\; 1 - \nu \; ;
\nonumber
\\
\chi \sim  \pi =  (\pi - \beta), \qquad {d^{2}g\over d\chi^{2}}-
{\nu(\nu + 1)\over \sin^{2} \chi}  \; g=0\,,
\nonumber
\\
 g = \sin^{B} (\pi - \beta)  \, , \qquad  B = \underline{1+ \nu}, \;\; - \nu \; .
\label{3.10b}
\end{eqnarray}

To simplify the problem  (\ref{3.9a}), it is convenient
to transform it to a new variable
$
e^{i\chi} = z$:
\begin{eqnarray}
\left [\;   {d^{2} \over dz^{2}} +  {1 \over z}  {d \over dz} - {4
\nu^{2}   \over  (z^{2} -1)^{2} } - { 2\nu (1+z^{2})  \over z
(z^{2}-1)^{2} } - {2m \epsilon \over z^{2}}   -  2i \alpha m \; {
z^{2} +1 \over z^{2} (z^{2}-1)}  \right  ] f = 0 \; .
\label{3.11a}
\end{eqnarray}

\noindent Analogously, eq. (\ref{3.9b}) gives
\begin{eqnarray}
[\;   {d^{2} \over dz^{2}} +  {1 \over z}  {d \over dz} - {4
\nu^{2}   \over  (z^{2} -1)^{2} }  +{ 2\nu (1+z^{2})  \over z
(z^{2}-1)^{2} } - {2m \epsilon \over z^{2}}   -  2i \alpha m \; {
z^{2} +1 \over z^{2} (z^{2}-1)}  \; ] f = 0 \; .
\label{3.11b}
\end{eqnarray}

\noindent They can be transformed into each other  via the formal
replacement  $\nu \Longrightarrow - \nu$. It suffices to examine
one of them. Let us consider eq. (\ref{3.11a}) near the  singular points
$z=\pm1,\; 0 $:

\vspace{3mm}

$
z = +1\;$,
\begin{eqnarray}
[\;   {d^{2} \over dz^{2}} +    {d \over dz} - {
\nu^{2}   \over  (z-1)^{2} } - { \nu   \over  (z-1)^{2} }
 \; ] f = 0 \; ,\qquad
  f = (z-1)^{A} , \qquad A  = \nu +1,\;\; - \nu \; ;
\nonumber
\end{eqnarray}

$
z = -1\;$,
\begin{eqnarray}
[\;   {d^{2} \over dz^{2}} -    {d \over dz} - {
\nu^{2}   \over  (z +1)^{2} } + { \nu   \over  (z+1)^{2} }
 \; ] f = 0
\;, \qquad  f = (z-1)^{B} , \qquad B = \nu,\;\; -\nu +1 \; ;
\nonumber
\end{eqnarray}

$
z = 0\;$,
\begin{eqnarray}
[\;   {d^{2} \over dz^{2}} +  {1 \over z}  {d
\over dz} - {2m \epsilon \over z^{2}}   +   \; {  2i \alpha m
\over z^{2} }  \; ] f = 0\; , \qquad f \sim z ^{\pm
\sqrt{2m\epsilon - 2i\alpha m }}\;.
\label{3.12}
\end{eqnarray}

It is convenient to translate all the  formulas to dimensionless
form. To this end, let energy unit be $ \hbar^{2} /  m
\rho^{2}$, where $\rho$  is a curvature radius, and two relevant
dimensionless parameters are ($q$ is an electron charge)
\begin{eqnarray}
E =
 \epsilon / {\hbar^{2} \over m \rho^{2}}  \; , \qquad e = {q^{2} \over \rho} /  {\hbar^{2} \over m \rho^{2}}\; ,
\nonumber
\end{eqnarray}

\noindent then instead of  (\ref{3.11a}), (\ref{3.11b}) we have
\begin{eqnarray}
\left [    {d^{2} \over dz^{2}} +  {1 \over z}  {d \over dz} - {4
\nu^{2}   \over  (z^{2} -1)^{2} } - { 2\nu (1+z^{2})  \over z
(z^{2}-1)^{2} } - {2E \over z^{2} }   -  2i e \; {  z^{2} +1 \over
z^{2} (z^{2}-1)}  \right ] f = 0 \; , \label{3.13a}
\\
\left [   {d^{2} \over dz^{2}} +  {1 \over z}  {d \over dz} - {4
\nu^{2}   \over  (z^{2} -1)^{2} } + { 2\nu (1+z^{2})  \over z
(z^{2}-1)^{2} } - {2E \over z^{2} }   -  2i e \; {  z^{2} +1 \over
z^{2} (z^{2}-1)}  \right  ] f = 0 \; . \label{3.13b}
\end{eqnarray}

For a time, for brevity let us use 'new'\hspace{1mm}  quantities
\begin{eqnarray}
2\nu  \; \Longrightarrow  \; \nu = 2j+1\; ,\qquad
 2E \; \Longrightarrow  \; E\;, \qquad  2ie \Longrightarrow e
\label{3.14}
\end{eqnarray}

\noindent which results
\begin{eqnarray}
\left [   {d^{2} \over dz^{2}} +  {1 \over z}  {d \over dz} - {
\nu^{2}   \over  (z^{2} -1)^{2} } - { \nu (1+z^{2})  \over z
(z^{2}-1)^{2} } - {E \over z^{2} }   -   e  {  z^{2} +1 \over
z^{2} (z^{2}-1)}  \right  ] f = 0 \; , \label{3.15a}
\\
\left [   {d^{2} \over dz^{2}} +  {1 \over z}  {d \over dz} - {
\nu^{2}   \over  (z^{2} -1)^{2} } + { \nu (1+z^{2})  \over z
(z^{2}-1)^{2} } - {E \over z^{2} }   -   e  {  z^{2} +1 \over
z^{2} (z^{2}-1)}  \right  ] f = 0 \; . \label{3.15b}
\end{eqnarray}

In eq.  (\ref{3.15a}), let us make all fractions simple ones
\begin{eqnarray}
\left [   {d^{2} \over dz^{2}} +  {1 \over z}
 {d \over dz} -{1\over 4}\,{\nu(\nu-2)\over (z+1)^{2}}+{e-E\over z^{2}}+{1\over 4}\,{\nu(\nu+2)
-4e\over z-1}- \right.
\nonumber
\\
\left. -{\nu\over z}-{1\over 4}\,{\nu(\nu+2)\over
(z-1)^{2}}+{1\over 4} \,{4e-\nu(\nu-2)\over z+1} \right ] f = 0
\;. \label{3.16}
\end{eqnarray}

\noindent With the substitution
$
f = z^{A} (z-1)^{B} (z+1)^{C} F(z)$ ,
 eq.  (\ref{3.16}) gives
\begin{eqnarray}
{d^{2}F\over dz^{2}}+\left[{2A+1\over z}+{2B\over z-1}+{2C\over
z+1}\right]{dF\over dz}+
\nonumber
\\
\left[{A^{2}+e-E\over z^{2}}+{B^{2}-B-1/4 \nu(\nu+2)\over
(z-1)^{2}}+{C^{2}-C-1/4\nu(\nu-2)\over (z+1)^{2}}+\right.
\nonumber
\\
+{BC+B+2AB+1/4[\nu(\nu+2)-4e]\over z-1}+{C+2AC-B-2AB-\nu\over z}+
\nonumber
\\
\left.+{-C-2AC-BC+1/4[4e-\nu(\nu-2)]\over z+1}\right]F=0\,.
\label{3.18}
\end{eqnarray}

\noindent At $A,\;B,\;C$ taken according to
\begin{eqnarray}
A^{2}+e-E=0\qquad\Rightarrow\qquad A=\pm \sqrt{E-e}\; ,
\nonumber
\\
B^{2}-B-1/4 \nu(\nu+2)=0\qquad\Rightarrow\qquad B=-{1\over
2}\,\nu\,,\;1+{1\over 2}\,\nu\; ,
\nonumber
\\
C^{2}-C-1/4\nu(\nu-2)=0\qquad\Rightarrow\qquad C={1\over
2}\,\nu\,,\;1-{1\over 2}\,\nu\; , \label{3.19}
\end{eqnarray}

\noindent eq.  (\ref{3.18}) becomes simpler
\begin{eqnarray}
{d^{2}F\over dz^{2}}+ \left[ {2A+1\over z} + {2B\over z-1}
+{2C\over z+1}  \right] \; {dF\over dz}+
\nonumber
\\
\left[  {C+2AC-B-2AB-\nu\over z}  +  {BC+B+2AB+1/4 [
\nu(\nu+2)-4e]\over z-1} + \right.
\nonumber
\\
\left.+{-C-2AC-BC+1/4[4e-\nu(\nu-2)]\over z+1}\right]F=0\,,
\label{3.20}
\end{eqnarray}

\noindent what is a   Hein equation for
 $G(p,\,q ; \, \alpha,\,\beta,\,\gamma,\,\delta ;  \, z)$
\begin{eqnarray}
{d^{2}F\over dz^{2}}+ \left[ \; {\gamma\over z}  +{\delta\over
z-1} + {\alpha+\beta-\delta-\gamma+1\over z-p}  \; \right]{dF\over
dz}+
\nonumber
\\
+\left[-{q\over p z}+{p\,\alpha\beta-q\over
p(p-1)(z-p)}+{-\alpha\beta+q\over (p-1)(z-1)}\right]F=0\, ,
\label{3.21}
\end{eqnarray}

\noindent when $p=-1$:
\begin{eqnarray}
{d^{2}F\over dz^{2}}+ \left ( \; {\gamma\over z}  +{\delta\over
z-1} + {\alpha+\beta-\delta-\gamma+1\over z-p}  \; \right
){dF\over dz}+
 \left ( {q\over   z}  + {\alpha\beta- q\over 2 (z-1)}  - {\alpha\beta + q\over 2 (z+1)} \right ) F=0\, .
\label{3.22}
\end{eqnarray}

Comparing (\ref{3.20}) with (\ref{3.22}), one finds expressions for parameters
\begin{eqnarray}
p=-1\, , \qquad q= (C  - B)(1+ 2A)-\nu\, ,
\nonumber
\\
\gamma=2A+1\, , \qquad \delta=2B\,; \qquad  ( \epsilon = 2C  )\; ;
\label{3.23}
\end{eqnarray}

\noindent and
\begin{eqnarray}
\alpha+\beta=2A+2B+2C\, ,
\nonumber
\\
\alpha\beta=B+C+2(AB+AC+BC)-2e+ \nu^{2}/2 \,,
\nonumber
\end{eqnarray}

\noindent that is
\begin{eqnarray}
\alpha=A+B+C-\sqrt{A^{2}+B^{2}+C^{2}-B-C+2e-\nu^{2}/2}\,,
\nonumber
\\
\beta=A+B+C+\sqrt{A^{2}+B^{2}+C^{2}-B-C+2e-\nu^{2}/2}\,.
\label{3.24}
\end{eqnarray}

Let
\begin{eqnarray}
 A=  + \sqrt{E-e}\,;
\qquad
 B=1+ \nu/ 2 \,;
\qquad
 C= \nu /2 \,,
\label{3.25a}
\end{eqnarray}

\noindent
 (positive values for   $B$ and  $C$ make solutions to be vanishing at the  points
  $z = \pm 1 \; (\chi = 0,\;  \pi $
), then
\begin{eqnarray}
\alpha=1+\nu+\sqrt{E-e}-\sqrt{E+e}\, , \qquad
\beta=1+\nu+\sqrt{E-e} +\sqrt{E+e}\,,
\nonumber
\end{eqnarray}

\noindent or (see  (\ref{3.14}))
\begin{eqnarray}
\alpha=2((j+1) +\sqrt{2E-2ie}-\sqrt{2E+2ie}\,,
\nonumber
\\
\beta= 2(j+1) +\sqrt{2E-2ie} +\sqrt{2E+2ie}\,. \label{3.25b}
\end{eqnarray}

Let us impose additional constraint (condition of polynomials)
\begin{eqnarray}
\beta = -2n \label{3.25c}
\end{eqnarray}

\noindent then a  quantization condition arises
\begin{eqnarray}
-  \sqrt{2E-2ie} - \sqrt{2E+2ie} =2 (n+j+1) \,,
\nonumber
\end{eqnarray}

\noindent  which after simple manipulation we   have arrived at a  formula  for energy levels
\begin{eqnarray}
E= - {e^{2} \over  2 (n+j+1)^{2} } +  {(n+j +1)^{2} \over 2} \; .
\label{3.26a}
\end{eqnarray}

It must be noted that the spectrum produced is very similar to
that for Schr\"{o}dinger's particle in Coulomb field; besides,
when  $e=$, it reduces to the exact formula for energy  levels for
a free particle in the space $S_{3}$.
With the use of (\ref{3.26a}), one can readily obtain rather simple
representation for all involved parameters. Indeed, (let $N =n +j
+1$; below we take the roots with negative real parts)
\begin{eqnarray}
\sqrt{ 2E  - 2ie} =
 \sqrt{ - {e^{2} \over N^{2}}  + N^{2}  -2ie } = \sqrt{ (N -{ie \over N} )^{2} } = - (N-{ie \over N}) \; ,
\nonumber
\\
\sqrt{ 2E  + 2ie} =
 \sqrt{ - {e^{2} \over N^{2}}  + N^{2}  +2ie } = \sqrt{ (N  + {ie \over N} )^{2} } = - (N +{ie \over N}) \; .
\label{3.26b}
\end{eqnarray}

\noindent Therefore,  $\alpha, \beta$  take the form
 \begin{eqnarray}
 \alpha = 2 (j+1) - (N-{ie \over N}) + (N +{ie \over N}) = 2(j+1) +{2ie \over n +j+1} \; ,
\nonumber
\\
 \beta = 2 (j+1) - (N-{ie \over N}) - (N +{ie \over N}) = - 2n \; .
\label{3.26c}
\end{eqnarray}

\subsection*{4.  Pauli equation for  Kepler problem, hyperbolic  space}

Let us start with  free radial equations (in which  the
rest energy is separated with the help of the
  formal replacement $\epsilon \Longrightarrow \epsilon + m$,
and the approximation $\epsilon = 2 m \approx 2m$ is used):
 \begin{eqnarray}
\delta =+1\;, \qquad ({d \over d \beta } \;+\; {\nu \over
\mbox{sinh}\; \beta }\;) \; f \; + \; 2 m \; g \; = \;0 \; ,\;\;
({d \over d \beta } \; - \;{\nu \over \mbox{sinh}\;
\beta }\;)\; g  -   \epsilon  \; f =\; 0     \; ;
\label{4.1a}
\\
\delta =-1\;, \qquad ({d \over d \beta } \;+\; {\nu \over
\mbox{sinh} \;  \beta }\;) \; f \; + \;
 \epsilon  \; g \; = \;0 \; ,\;\;
 ({d \over d \beta} \; - \;{\nu \over
\mbox{sinh}\; \beta }\;)\; g  \;- \;  2m \; f =\; 0     \;.
\label{4.1b}
\end{eqnarray}

\noindent In each case one gets
 a radial Pauli equation for a  big  2-component:
 \begin{eqnarray}
\delta =+1\;, \qquad f >> g \; ,\qquad  {d^{2}f\over d \beta
^{2}}-\left({\nu(\nu+ \mbox{ch} \; \beta ) \over \mbox{sinh}^{2}
\beta }-2\epsilon m\right) \; f=0\; ;
\label{4.2a}
\\
\delta =-1\;, \qquad g >> f \; , \qquad  {d^{2}g\over d\beta^{2}}-
\left({\nu(\nu- \mbox{ch}\; \beta)\over \mbox{sinh}^{2}
\beta}-2\epsilon m\right) \; g =0\;.
\label{4.2b}
\end{eqnarray}

\noindent Corresponding  wave functions for states with different
parity are of the form
 \begin{eqnarray}
\psi _{jm,\delta=+1}={ e^{i\epsilon t} \over \mbox{sinh}\;
\beta}\; \left |
\begin{array}{c}
f(\beta) \; D_{-1/2}\\
f(\beta) \; D_{+1/2}
\end{array} \right |\;,
\qquad \psi _{jm,\delta=-1}={ e^{i\epsilon t} \over \mbox{sinh}\;
\beta}\; \left |
\begin{array}{c}
ig\;(\beta) \; D_{-1/2}\\
-ig\;(\beta) \; D_{+1/2}
\end{array} \right |\;.
\label{4.3}
\end{eqnarray}

Now let us consider the Coulomb field. It is enough to make a
formal replacement  in  (\ref{4.1a}), (\ref{4.1b})
 \begin{eqnarray}
\delta =+1\;, \qquad ({d \over d \beta } \;+\; {\nu \over
\mbox{sinh}\; \beta }\;) \; f \; + \; 2 m \; g \; = \;0 \; ,
\nonumber
\\
\qquad \qquad \qquad ({d \over d \beta } \; - \;{\nu \over \mbox{sinh}
\;\beta }\;)\; g  -   (\epsilon  + {\alpha \over  \mbox{tanh}\;
\beta} ) \; f =\; 0     \; ; \label{4.4a}
\\
\delta =-1\;, \qquad ({d \over d \beta } \;+\; {\nu \over
\mbox{sinh}\;  \beta }\;) \; f \; + \;
 (\epsilon   + {\alpha \over  \mbox{tanh}\; \beta}) \; g \; = \;0 \; ,
\nonumber
\\
\qquad \qquad \qquad ({d \over d \beta } \; - \;{\nu \over
\mbox{sinh} \;\beta }\;)\; g  \;- \;  2m \; f =\; 0     \; .
\label{4.4b}
\end{eqnarray}

For each value of parity one obtains its differential equation
\begin{eqnarray}
{d^{2}f\over d\beta^{2}}- \left({\nu(\nu+ \mbox{ch} \beta )\over
\mbox{sinh}^{2} \beta}- 2\epsilon m-{2 m \alpha \over \mbox{tanh}
\; \beta }\right)  f=0\,, \label{4.5a}
\\
{d^{2}g\over d\beta^{2}}-\left({\nu(\nu- \mbox{ch}\; \beta )\over
\mbox{sinh}^{2} \beta }-2\epsilon m-{2 m \alpha \over  \mbox{tanh}
\beta}\right)  g =0 \; . \label{4.5b}
\end{eqnarray}

Let us study eq. (\ref{4.5a}). To simplify the problem it is convenient
to transform it to a new variable $e^{\beta} = z$. As in the
spherical space we will use to dimensionless variables and use the
notation:
\begin{eqnarray}
 2\nu \Longrightarrow \nu = 2j +1 \; , \qquad
 2E \; \Longrightarrow  \; E\;, \qquad  2e \Longrightarrow e \; ,
\label{4.6}
\end{eqnarray}

\noindent then
\begin{eqnarray}
\left [   {d^{2} \over dz^{2}} +  {1 \over z}  {d \over dz} -
{1\over 4}{\nu(\nu-2)\over (z+1)^{2}}+ {E - e  \over
z^{2}}+{1\over 4}{\nu(\nu+2) + 4e \over z-1}- \right.
\nonumber
\\
\left. -{\nu\over z}-{1\over 4}\,{\nu(\nu+2)\over
(z-1)^{2}}-{1\over 4}\,{4 e + \nu(\nu-2)\over z+1} \right ] f =
0\,.
\label{4.7}
\end{eqnarray}

\noindent
With  the substitution $ f = z^{A}\, (z-1)^{B}\, (z+1)^{C}\, F(z)$,
(\ref{4.7}) gives
\begin{eqnarray}
{d^{2}F\over dz^{2}}+\left[{2A+1\over z}+{2B\over z-1}+{2C\over
z+1}\right]{dF\over dz}+
\nonumber
\\
+\left[{A^{2}+ E - e \over z^{2}}+{B^{2}-B-1/4\nu(\nu+2)\over
(z-1)^{2}}+{C^{2}-C-1/4\nu(\nu-2)\over (z+1)^{2}}+\right.
\nonumber
\\
+{BC+B+2AB+1/4\nu(\nu+2) + e \over z-1}+{C+2AC-B-2AB-\nu\over z}+
\nonumber
\\
\left.+{-C-2AC-BC- e -1/4\nu(\nu-2)\over z+1}\right]F=0\,.
\label{4.8}
\end{eqnarray}

\noindent At $A,\;B,\;C$ taken according to
\begin{eqnarray}
A^{2}+ E -e  =0\qquad\Rightarrow\qquad A=\pm \sqrt{e -E}\,;
\nonumber
\\
B^{2}-B-1/4\nu(\nu+2)=0\qquad\Rightarrow\qquad B=-{1\over
2}\,\nu\,,\;1+{1\over 2}\,\nu\,;
\nonumber
\\
C^{2}-C-1/4\nu(\nu-2)=0\qquad\Rightarrow\qquad C={1\over
2}\,\nu\,,\;1-{1\over 2}\,\nu\,, \label{4.9}
\end{eqnarray}

\noindent eq.  (\ref{4.8}) becomes simpler
\begin{eqnarray}
{d^{2}F\over dz^{2}}+\left[{2A+1\over z}+{2B\over z-1}+{2C\over
z+1}\right]{dF\over dz}+
\nonumber
\\
+\left[{BC+B+2AB+1/4\nu(\nu+2) +e \over z-1}+{C+2AC-B-2AB-\nu\over
z}+\right.
\nonumber
\\
\left.+{-C-2AC-BC- e -1/4\nu(\nu-2)\over z+1}\right]F=0\,,
\label{4.10}
\end{eqnarray}

\noindent what is a   Heun equation for
 $G(p,\,q ; \, \alpha,\,\beta,\,\gamma,\,\delta ;  \, z)$
\begin{eqnarray}
p=-1\, , \qquad q=C+2AC-B-2AB-\nu\,;
\nonumber
\\
\gamma=2A+1\, , \qquad \delta=2B\, , \label{4.11}
\end{eqnarray}

\noindent and
\begin{eqnarray}
\alpha+\beta=2A+2B+2C\,;
\nonumber
\\
\alpha\beta=B+C+2(AB+AC+BC)+{1\over 2}\nu^{2}+2e \,;
\nonumber
\end{eqnarray}

\noindent that is
\begin{eqnarray}
\alpha=A+B+C-\sqrt{A^{2}+B^{2}+C^{2}-B-C-1/2\nu^{2}- 2e}\,,
\nonumber
\\
\beta=A+B+C+\sqrt{A^{2}+B^{2}+C^{2}-B-C-1/2\nu^{2}-2e}\,.
\label{4.12}
\end{eqnarray}

Let
\begin{eqnarray}
 A= - \sqrt{e-E}\,;
\qquad
 B=1+{1\over 2}\,\nu\,;
\qquad
 C={1\over 2}\,\nu\, ;
\label{4.13}
\end{eqnarray}

\noindent the negative value of $A$ ensures  vanishing the
function  at the infinity $\chi \rightarrow +\infty$. The positive
value of $B$ ensures vanishing of the  function  in the origin.
Then
\begin{eqnarray}
\alpha=1+\nu - \sqrt{e-E} - \sqrt{-e -E}\, , \qquad \beta=1+\nu -
\sqrt{e-E} + \sqrt{-e - E}\, ,
\nonumber
\end{eqnarray}

\noindent or remembering about (\ref{4.6})
\begin{eqnarray}
\alpha=2(j+1) -\sqrt{2e-2E}- \sqrt{-2E- 2e}\,,
\nonumber
\\
\beta= 2(j+1) - \sqrt{2e-2E} +\sqrt{-2E-2e}\, . \label{4.14}
\end{eqnarray}

\noindent Imposing  additional constraint (condition for polynomial solutions)
\begin{eqnarray}
\alpha = -2n \; ; \label{4.15}
\end{eqnarray}

\noindent we obtain
\begin{eqnarray}
\sqrt{2e-2E} +   \sqrt{-2E-2e}  =2 (n+j+1) \, ,
\nonumber
\end{eqnarray}

\noindent which after simple manipulation gives
 a formula for energy levels
\begin{eqnarray}
E= - {e^{2} \over  2 (n+j+1)^{2} } -  {(n+j +1)^{2} \over 2} \; .
\label{4.16}
\end{eqnarray}

With the use of (\ref{4.16}), one can readily obtain rather simple
representation for involved parameters
\begin{eqnarray}
\alpha =  2(j+1) - N - {e \over N} - N + {e \over N} = -2n \,,
\nonumber
\\
\beta=  2(j+1)- N - {e \over N} + N - {e \over N}= 2(j+1) - {2e
\over n+j+1} \, . \label{4.17}
\end{eqnarray}

Author plans  to consider relativistic Coulomb problem on the base of the Dirac equation
in space of constant curvature. Such a problem turns to be much  more complicated -- it reduces to
a second order differential equation with 6 singular points. With special mathematical manipulations
we can reduce the problem to a differential equation with 5 singular points, however
it still remains  very  difficult  mathematical task.

\subsection*{Acknowledgement}

Author  is grateful  to V.M. Red'kov for  moral support and advice.

\end{document}